\documentclass[a4paper]{article}
\usepackage{graphicx}

\begin{document}

\thispagestyle{empty}

\noindent
\textbf{Preprint of:}\\
T. A. Nieminen,
N. R. Heckenberg and
H. Rubinsztein-Dunlop\\
``Optical measurement of microscopic torques'' \\
\textit{Journal of Modern Optics} \textbf{48}, 405-413 (2001)\\
Changes: equation (\ref{measured_power}) corrected.

\begin{center}

\Large {\bf
Optical measurement of microscopic torques \\
}

\vspace{1pc}
\large

T. A. Nieminen,
N. R. Heckenberg and
H. Rubinsztein-Dunlop

\vspace{1pc}
\small
\textit{
Centre for Laser Science, Department of Physics \\
The University of Queensland, Brisbane, QLD 4072, Australia.\\
tel: +61-7-3365 3405, fax: +61-7-3365 1242, \\
e-mail: timo@physics.uq.edu.au
}
\vspace{1pc}
\end{center}

\begin{abstract}
In recent years there has been an explosive development of interest in the measurement
of forces at the microscopic level, such as within living cells
\cite{quake1997,dai1999,sleep1999}, as well as the
properties of fluids and suspensions on this scale \cite{grier1997},
using optically trapped
particles as probes. The next step would be to measure torques and associated
rotational motion \cite{ryu2000}.
This would allow measurement on very small scales since no
translational motion is needed. It could also provide an absolute measurement of the
forces holding a stationary non-rotating particle in place. The laser-induced
torque acting on an optically trapped microscopic birefringent particle \cite{friese1998} can
be used for these measurements. Here we present a new method for simple, robust,
accurate, simultaneous measurement of the rotation speed of a laser trapped
birefringent particle, and the optical torque acting on it, by measuring the
change in angular momentum of the light from passing through the particle.
This method does not depend on the size
or shape of the particle or the laser beam geometry, nor does it depend
on the properties of the surrounding medium.
This could allow accurate measurement of viscosity on a microscopic scale.
\end{abstract}

\section{Introduction}

Optical torques have been measured previously; two methods have been used.
Firstly, if a particle with known birefringent properties has a simple and
accurately known size and shape, ideally a flat disc, the torque can be
calculated from the beam power \cite{friese1998}. However, particles of more
complex shapes will often be used in experiments or encountered in samples;
for example, spherical
particles are ideal for making measurements of viscosity. Secondly, torques
have been determined by measuring the rotation speed in a medium of known
viscosity \cite{friese1998,luo2000}. This method cannot be used if the
aim is to measure an unknown viscosity. This method would also fail if there
were other torques acting on the particle, or if the viscous drag is affected
by nearby walls or other particles, or if the particle is not rotating.
Previous methods for measuring rotation speeds, based on the periodic
variation of backscattered light \cite{friese1996}, can also have problems.
Very regular or
rotationally symmetric particles provide insufficient variation or variation
at an increased frequency. Consideration of the basic physical processes
giving rise to the torque gives a new method for measuring the torque and
rotation speed that overcomes all of these problems.

Since optical torques and forces are very small, microscopic particles are ideal
for the observation and application of optical torques and rotation. Such
microscopic particles will typically be confined within a laser trap. Strongly
focussed laser light incident on a transparent particle, usually in a liquid medium,
will produce a gradient force acting on the particle towards the region of highest
irradiance. If this gradient force near the focus is stronger than scattering and
absorption forces, the particle will be trapped at the beam focus, where the irradiance
is highest. This technique of three-dimensional confinement and manipulation is
called laser micro-manipulation, trapping, or optical tweezers.

\section{Polarised beams}

A monochromatic laser beam can be written as a plane wave in terms of two orthogonal
components:
\begin{equation}
\mathbf{E} = ( E_x \hat{\mathbf{x}} + E_y \hat{\mathbf{y}} ) \exp(i k z - i \omega t)
\label{planecomponents}
\end{equation}
where the beam is propagating in the $z$-direction. In general, the amplitudes
$E_x$ and $E_y$ are complex in order to account for the phases of the components.

There are two cases of special interest.
The first is when the phase angle between the complex
amplitudes $E_x$ and $E_y$ is equal to 0 or $\pi$, in which case the total electric field
always lies in a single plane resulting in a beam which is linearly polarised.
The direction of the
$x$-axis can be chosen to coincide with the plane of polarisation,
so the beam can be written as
\begin{equation}
\mathbf{E} = E_p \hat{\mathbf{x}} \exp(i k z - i \omega t)
\label{plane}
\end{equation}
where $E_p$ is the complex amplitude of the linearly polarised light

The second special case is when the phase angle is $\pm \pi / 2$, and $|E_x| = |E_y|$.
In this case, $E_y = E_x \exp i \theta$, with $\theta = \pm \pi / 2$.
The total electric field has a constant magnitude, with the direction varying with the
optical frequency $\omega$ so that the beam is circularly polarised.
When $\theta = + \pi / 2$,
the electric field has a positive, or right-handed, helicity. Such a beam is here
called left circularly polarised. When $\theta = -\pi / 2$, the beam has negative
helicity and is called right circularly polarised.
A circularly polarised beam can always be written
as
\begin{equation}
\mathbf{E} = ( E_c \hat{\mathbf{x}} \pm i E_c \hat{\mathbf{y}} ) \exp(i k z - i \omega t)
\label{circular}
\end{equation}
with the sign depending on whether the beam is left or right circularly
polarised, and $E_c$ is the complex amplitude.

In general, however, the phase angle $\theta$ will have a value between these limiting
values, or even if $\theta = \pm \pi / 2$, $|E_x| \neq |E_y|$. In these cases, the beam is
elliptically polarised, and the electric field vector $\mathbf{E}$ traces out an ellipse
during each optical period.

Recognising that we can rewrite equation~(\ref{circular}) for a circularly polarised beam as
\begin{equation}
\mathbf{E} = E_c (\hat{\mathbf{x}} \pm i \hat{\mathbf{y}}) \exp(i k z - i \omega t),
\end{equation}
we see that any beam can be represented as a sum of two circularly polarised
components using the (complex) orthogonal basis vectors
\begin{eqnarray}
\mathbf{\hat{e}}_L & = &  \frac{1}{\sqrt{2}} ( \hat{\mathbf{x}} + i \hat{\mathbf{y}} )
\nonumber \\
\mathbf{\hat{e}}_R & = & \frac{1}{\sqrt{2}} ( \hat{\mathbf{x}} - i \hat{\mathbf{y}} )
\end{eqnarray}
as
\begin{equation}
\mathbf{E} = ( E_L \mathbf{\hat{e}}_L + E_R \mathbf{\hat{e}}_R ) \exp(i k z - i \omega t).
\label{circularcomponents}
\end{equation}
The amplitudes of the left and right circular components can be found from the $x$ and
$y$ amplitudes in the linear orthogonal representation (equation~(\ref{planecomponents})):
\begin{eqnarray}
E_L & = & \frac{1}{\sqrt{2}} ( E_x - i E_y )
\nonumber \\
E_R & = & \frac{1}{\sqrt{2}} ( E_x + i E_y )
\end{eqnarray}

When $|E_L|=|E_R|$, the beam is linearly polarised, with the plane of polarisation given by
the phase angle between the complex amplitudes $E_L$ and $E_R$. If $E_L=0$ (and $E_R\ne0$),
the beam is right circularly polarised, and left circularly polarised if $E_R=0$.

The time-averaged irradiance is given by \cite{hecht}
\begin{equation}
I = \frac{c \epsilon_0 E_L^\star E_L}{2} + \frac{c \epsilon_0 E_R^\star E_R}{2} = I_L + I_R.
\end{equation}

Although we have only considered the beam as a classical EM wave so far, the fact that
the angular momentum of left and right circularly polarised photons is $\pm \hbar$ can
be used to simply find the angular momentum of the beam.
Since the energy of a photon is $\hbar \omega$, the photon flux per unit area is
\begin{equation}
N = \frac{I}{\hbar \omega} = \frac{I_L}{\hbar \omega} + \frac{I_R}{\hbar \omega},
\end{equation}
giving an angular momentum flux per unit area of
\begin{equation}
L_z = ( I_L - I_R ) / \omega.
\end{equation}
Thus, the beam can be considered to have a net circularly polarised component
with a power of $|I_L-I_R|$
which contributes to the angular momentum of the beam, and a linearly polarised
component of $I - |I_L-I_R| = 2 \min(I_L,I_R)$ which does not contribute to the
angular momentum of the beam. We can define a coefficient of circular polarisation
$\sigma_z$ by
\begin{equation}
\sigma_z = ( I_L - I_R ) / I,
\end{equation}
and write the angular momentum flux density of the beam as
\begin{equation}
L_z = \sigma_z I / \omega.
\end{equation}

When the irradiance is integrated across the whole beam, the total power can be
obtained and will be given by
$P = P_L + P_R = \int I_L dA + \int I_R dA$. A suitable average coefficient
of circular polarisation can be defined by
\begin{equation}
\sigma_z = ( P_L - P_R ) / P,
\end{equation}
with the resulting total angular momentum flux of the beam being
\begin{equation}
L_z = \sigma_z P / \omega.
\end{equation}

\section{Optical torque}

If the beam passes through some birefringent material, the polarisation will be affected.
In general, $\sigma_z$ will change. The incident beam will have
an initial coefficient of circular polarisation
$\sigma_{zin}$, and will have an emergent
polarisation described by $\sigma_{zout}$. Thus the angular momentum of the beam will change,
and a reaction torque on the birefringent material will result.
The reaction torque is equal to the change in the angular momentum flux:
\begin{equation}
\tau = ( \sigma_{zin} - \sigma_{zout} ) P / \omega
\label{torque}
\end{equation}
assuming that absorption and reflection can be ignored. Equation~(\ref{torque}) is general,
and can always be used to find the torque if the coefficients of circular polarisation
of the incident and outgoing beams are known or can be found.

Although equation~(\ref{torque}) applies in general, it is instructive to carry through
a detailed calculation for a simple case: a uniform sheet of birefringent material,
for example calcite.

A uniaxial birefringent material such as calcite can be described by two
refractive indices: an \emph{ordinary refractive index} $n_o$ for electric
fields normal to the optic axis, and an \emph{extraordinary refractive index} $n_e$
for electric fields parallel to the optic axis. For calcite, $n_o = 1.66$
and $n_e = 1.49$.
Consider a thickness $d$ of uniaxial birefringent material with the optic axis
in the $xy$ plane, at an angle of $\theta$ to the $x$-axis. The front face of the
material is at $z = z_0$, and the rear face is at $z = z_0 + d$.
If the electric field of the incident beam at the front surface of the material
is given by equation~(\ref{planecomponents}),
we can express this in terms of unit vectors $\hat{\mathbf{i}}$ and
$\hat{\mathbf{j}}$ parallel to and normal to the optic axis:
\begin{eqnarray}
\mathbf{E} & = & [ ( E_x \cos \theta + E_y \sin \theta ) \hat{\mathbf{i}}
                 + ( - E_x \sin \theta + E_y \cos \theta ) \hat{\mathbf{j}} ]
\nonumber \\ & &
                 \times \exp(i k z_0 - i \omega t).
\end{eqnarray}
In terms of circular components, this gives
\begin{eqnarray}
\mathbf{E} & = & \frac{1}{\sqrt{2}} [ ( E_x - i E_y ) \exp(i\theta) \mathbf{\hat{e}}_L'
                                      + ( E_x + i E_y ) \exp(-i\theta) \mathbf{\hat{e}}_R' ]
\nonumber \\ & &
             \times \exp(i k z_0 - i \omega t)
\end{eqnarray}
where $\mathbf{\hat{e}}_L' = \frac{1}{\sqrt{2}} (\hat{\mathbf{i}}+i\hat{\mathbf{j}})$
and $\mathbf{\hat{e}}_R' = \frac{1}{\sqrt{2}} (\hat{\mathbf{i}}-i\hat{\mathbf{j}})$
The coefficient of circular polarisation is given by
\begin{eqnarray}
\sigma_{zin} & = & \frac { E_L^\star E_L - E_R^\star E_R } { E_L^\star E_L + E_R^\star E_R }
\nonumber \\
             & = & \frac { i ( E_x E_y^\star - E_x^\star E_y ) } { E_x^\star E_x + E_y^\star E_y }.
\end{eqnarray}
After passing through the thickness $d$, the field will be
\begin{eqnarray}
\mathbf{E} & = & [ ( E_x \cos \theta + E_y \sin \theta ) \exp(ikdn_e) \hat{\mathbf{i}}
\nonumber \\ & &
                   + ( - E_x \sin \theta + E_y \cos \theta ) \exp(ikdn_o)  \hat{\mathbf{j}} ]
                   \exp(i k z_0 - i \omega t),
\end{eqnarray}
which we can express in terms of circular components
\begin{eqnarray}
E_L & = & \frac{1}{\sqrt{2}} [ ( E_x \cos\theta + E_y \sin\theta ) \exp(i k d n_e)
\nonumber \\
    &   &                      - i ( - E_x \cos\theta + E_y \sin\theta ) \exp(i k d n_o) ]
\nonumber \\
E_R & = & \frac{1}{\sqrt{2}} [ ( E_x \cos\theta + E_y \sin\theta ) \exp(i k d n_e)
\nonumber \\
    &   &                      + i ( - E_x \cos\theta + E_y \sin\theta ) \exp(i k d n_o) ]
\end{eqnarray}

We define the convenient notation
\begin{equation}
\Delta = k d ( n_o - n_e )
\label{deltadefn}
\end{equation}

The coefficient of circular polarisation of the emergent light is
\begin{eqnarray}
\sigma_{zout} & = & [ i\cos\Delta ( E_x E_y^\star - E_x^\star E_y )
             \nonumber \\ & &
             - \sin\Delta
             \{ ( E_x^\star E_x - E_y^\star E_y ) \sin2\theta
             - ( E_x E_y^\star + E_x^\star E_y ) \cos2\theta \} ]
             \nonumber \\ & &
             / ( E_x^\star E_x + E_y^\star E_y )
\label{sigmazout}
\end{eqnarray}
giving a torque per unit area of
\begin{eqnarray}
\tau & = & \frac{c\epsilon_0}{2\omega}
      [ i ( E_x E_y^\star - E_x^\star E_y ) ( 1 - \cos\Delta )
      \nonumber \\ & &
      + \sin\Delta
      \{ ( E_x^\star E_x - E_y^\star E_y ) \sin2\theta
      - ( E_x E_y^\star + E_x^\star E_y ) \cos2\theta \} ]
\end{eqnarray}
If the incident light is linearly polarised ($E_y = 0$), the torque is
\begin{equation}
\tau = \frac{c\epsilon_0}{2\omega} \sin\Delta E_0^\star E_0 \sin2\theta,
\end{equation}
which acts to align the slow axis of the particle with the plane of polarisation
if $n_o > n_e$, or normal to the plane of polarisation if $n_e > n_o$.
If the incident light is left circularly polarised ($E_y = iE_x$), the torque is
\begin{equation}
\tau = \frac{c\epsilon_0}{\omega} E_0^\star E_0 ( 1 - \cos\Delta )
\end{equation}
which is independent of the orientation of the birefringent material.

If the birefringent material is of a uniform thickness, the total torque
can be simply
calculated from this \cite{friese1998}. In general, a laser trapped
birefringent particle will have a varying thickness, and direct calculation
of the torque will not be feasible. Also, if the orientation of the birefringent
particle is different, so the the optic axis does not lie in the $xy$ plane, the
calculation will be further complicated. Equation~(\ref{torque}), however,
is general, and will still apply, and the torque acting on the particle can
be deduced from the change in polarisation of the light.

\section{Optical torque measurement}

Consider a circularly polarised laser beam used to trap a microscopic particle
composed of a uniaxial birefringent material such
as calcite or a suitable polymer. If the optical torque is
large enough to overcome forces holding the particle in place, the particle will
rotate at a speed determined by the equilibrium between the optical torque and other
forces such as viscous drag. In this way, a probe particle can be used to measure
viscosity on a microscopic scale. If the particle does not rotate, the optical
torque can be used to determine the torque due to static forces acting on the particle.

The maximum torque and rotation rate will occur when the incident beam is
completely circularly polarised (ie $\sigma_{zin} = 1$). The torque in this case will also
be constant as well as maximal \cite{friese1998}, and we will only consider this case
here. The torque $\tau$ acting on the trapped particle is given in equation~(\ref{torque})
by the difference between the
incident and outgoing angular momentum fluxes, and in this case, assuming no
reflection or absorption, is
\begin{equation}
\tau = ( 1 - \sigma_{zout} ) P / \omega.
\end{equation}
Measurement of the outgoing polarisation $\sigma_{zout}$ and beam power $P$ gives an
absolute measurement of the torque, which does not depend on the mechanical
properties of the surrounding medium or the particular size or shape of the particle
or laser beam.

We can also note that the plane of polarisation of the linearly polarised component of the
outgoing beam exiting a rotating birefringent particle will be rotating at the same
rotation rate $\Omega$
as the particle. If the outgoing beam is a (rotating) purely plane-polarised beam, as
would occur if the particle acted as a quarter-wave plate, rotating at $\Omega$, and of
power $P$, and is passed through a linear polariser, the measured power $P_m$ will be
$P_m = (1 + \cos 2 \Omega t) P/2$ (with variation at a frequency of $2 \Omega$ since a
rotation of $180^\circ$ rotates the plane of polarisation onto itself) \cite{bagini1994}.
By measuring this power, the rotation rate $\Omega$ of the trapped particle can be
simply determined. This will still be the case for an elliptically polarised beam, as the
same variation at a frequency of $2 \Omega$ will be observed. The angular momentum
associated with this rotation of the plane of polarisation will be negligible as
$\Omega \ll \omega$.

In the general case, there will be an elliptically polarised outgoing beam, consisting
of both plane and circularly polarised components.
The measured power
$P_m$ after the outgoing beam passes through a linear polariser acting as an analyser
will be
\begin{equation}
P_m = \{ 1 + ( 1 - \sigma_{zout}^2 )^{1/2} \cos 2 \Omega t \} P / 2.
\label{measured_power}
\end{equation}
Measurement of the variation of the transmitted power therefore allows the
determination of the rotation period of the trapped particle, and the degree
of (but not the direction of) circular polarisation. The result of a measurement
of this type will be as shown in figure~1. This measurement is an average
over the beam, and it is not important whether or not the entire beam passes through
the particle. In the case where the particle is not rotating, due to some restraining
torque, the plane of polarisation of the transmitted light will not be rotating.
The degree of circular polarisation can be measured in this case by rotating the
linear polariser, which will give the same result where $\Omega$ is the rotation
rate of the polariser relative to the particle. The orientation of the particle
can also be determined from the position of the measured power maxima.

\begin{figure}[htb]
\begin{center}
\includegraphics[width=\textwidth]{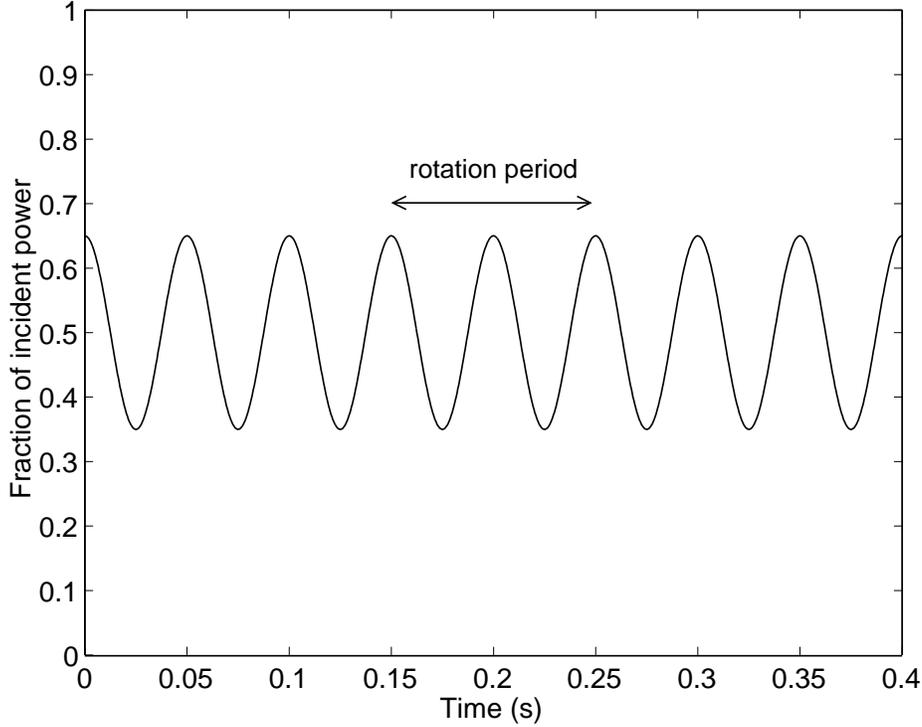}
\end{center}
\caption{The power which would be measured through a plane polariser after the beam has
passed through a birefringent particle is shown for a beam with $\sigma_{zin}=+1$,
$\sigma_{zout} = +0.7$ and particle rotation frequency $\Omega = 10$~Hz. The mean
power measured through the polariser is half of the power incident on the particle.
The frequency of the variation is two times the rotation rate $\Omega$ of the
particle. The optical torque can be found from the amplitude of the variation and
the measured power once the direction of the transmitted polarisation is known. In this
case, for a 100~mW trapping beam of wavelength 1,064~nm, an optical torque of
16.9~pN$\cdot$$\mu$m is being exerted.}
\end{figure}

In many cases, the direction of the transmitted polarisation will be known
beforehand---such
as when the particle is insufficiently thick to change the direction of
polarisation (note that a calcite particle approximately $3 \mu m$ thick is a
$\lambda / 2$ plate for 1064~nm light), or when the particle is small and the
outgoing light is dominated by light that did not pass through the particle and has
not changed in polarisation. If necessary, the direction of circular polarisation
can be measured simply by placing a reversed circular polariser (eg a quarter-wave
plate followed by a linear polariser appropriately oriented) in the beam path instead
of a linear polariser. In the case where the trapping beam has a right-handed helicity
(left circularly polarised, with $\sigma_{zin} = +1$), the light emergent from the
particle can be described in terms of left and right circularly polarised components
$P_L$ and $P_R$, where $P_L = ( 1 + \sigma_{zout} ) P / 2$ and
$P_R = ( 1 - \sigma_{zout} ) P / 2$. If the output beam is predominantly left circularly
polarised, $\sigma_{zout} > 0$, and $P_L > P_R$. A right circularly polarised beam
has $\sigma_{zout} < 0$, and $P_R > P_L$. It is only necessary to determine which of these
two components is larger, rather than to measure each one individually since
$|\sigma_{zout}|$ is already known. In this way, the direction of circular polarisation
can be determined, and $\sigma_{zout}$ as opposed to merely $|\sigma_{zout}|$ can be
found. Once $\sigma_{zout}$ is known, the optical torque acting on the particle
can be found using equation~(\ref{torque}). A measurement of this type will be as shown
in figure 2.

\begin{figure}[htb]
\begin{center}
\includegraphics[width=\textwidth]{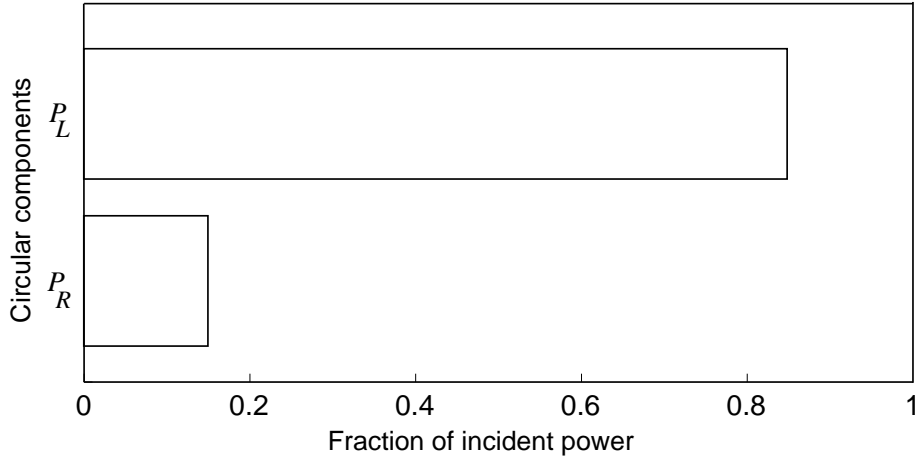}
\end{center}
\caption{The two circularly polarised components of the transmitteed light can be measured
to determine the direction of circular polarisation. $P_L$ is the power of the left
circularly polarised component, and $P_R$ is the power of the right circularly
polarised component.}
\end{figure}

It should be noted that this technique is robust. It is not necessary to measure the
power of the entire transmitted beam; it is sufficient to measure the portion of the
beam that has passed through the trapped particle.
Similarly, reflections are not
likely to cause significant error. Some of the incident beam will be reflected from
the trapped particle; the reflection will depend on the angle of incidence and the
refractive indices of the particle and the surrounding medium.
For example, for calcite trapped in water, the Fresnel amplitude coefficients for
reflection at normal incidence are \linebreak
$(n_{water}-n_{calcite})/(n_{water}+n_{calcite})$,
which gives reflected amplitudes of $-0.057E_x$ and $-0.11iE_x$ for linearly
polarised components parallel to and normal to the optic axis respectively.
In terms of circular components, this becomes $E_L=-0.02E_{L0}$ and
$E_R=-0.08E_{L0}$, showing that the torque due to backreflected light
will be less than 0.6\% of the available torque.
Therefore, the reflected light will not cause any significant error.

\section{Conclusion}

A simple method of measuring the rotation speed and the optical torque applied
to a laser trapped birefringent particle has been described. This method can be
used even if the viscosity of the medium in which trapping is performed is unknown,
and provides a means to measure this viscosity. Thus, this method is suitable for
employment in a micro-rheometer, which could be simply constructed by trapping a
birefringent probe particle in the fluid of interest. A suitable test particle would
be a small fragment of calcite, the exact shape not being critical at the very low
Reynolds numbers encountered in these cases, or a more ideal shape could be fabricated
from a birefringent polymer \cite{higurashi1997}. As the optical torque can be
controlled by varying the power, the probe particle rotation speed can be varied,
allowing, for example, the investigation of non-linear properties of the fluid.
Wall effects and the small-scale behaviour of polymer and colloidal suspensions
could be investigated, or even the rheological properties of intracellular
fluids or membranes in vivo.

\section*{Acknowledgements}

This work has been supported by the Australian Research Council. The authors would like
to thank Drs Friese and Bishop for valuable discussions.

\end{document}